\documentclass[12pt]{article}
\usepackage{epsf,epsfig}
\usepackage{amsmath,amsfonts,amssymb}
\usepackage{cite}
\usepackage{breqn}

\newcommand{\be}{\begin{equation}}
\newcommand{\ee}{\end{equation}}
\newcommand{\ben}{\begin{eqnarray}\displaystyle}
\newcommand{\een}{\end{eqnarray}}

\newcommand{\bea}[1]{\begin{eqnarray}\label{#1} }
\newcommand{\eea}{\end{eqnarray}}

\newcommand{\refb}[1]{(\ref{#1})}

\newcommand{\sectiono}[1]{\section{#1}}
\newcommand{\subsectiono}[1]{\subsection{#1}}

\def\boxempty{{\,\lower0.9pt\vbox{\hrule \hbox{\vrule height 0.25 cm
\hskip 0.25 cm \vrule height 0.25 cm}\hrule}\,}}

\def\one{{\hbox{ 1\kern-.8mm l}}}
\def\zero{{\hbox{ 0\kern-1.5mm 0}}}

\begin{document}
\begin{titlepage}
\thispagestyle{empty}

\title{
{\Large\bf Travelling Front of a Decaying Brane}\\
{\Large\bf in String Field Theory}
}

\author{
{\large\bf Debashis Ghoshal
and
Preeda Patcharamaneepakorn}\\
{}\\
{\large\it School of Physical Sciences}\\
{\large\it Jawaharlal Nehru University}\\
{\large\it New Delhi 110067, India}\\
{}\\
{\tt dghoshal@mail.jnu.ac.in}\\
{\tt preeda.pat@gmail.com}\\
{}\\
}

\bigskip\bigskip

\date{%
%
\bigskip\bigskip\bigskip\bigskip
\begin{quote}
\centerline{\large\bf Abstract:}
{\small
We consider the inhomogeneous decay of an unstable D-brane of bosonic string
theory in a linear dilaton background in a light-cone frame. At the lowest level, the
dynamical equation that describes this process is a generalisation (that includes
nonlocality and time delay) of a reaction-diffusion equation studied by Fisher (and
others). We argue that the equation of motion of the cubic open string field theory
is satisfied at least to the second order when we start with this `Fisher deformation',
a marginal operator which has a simple pole term in its OPE. We also compute the
one-point functions of closed string operators on the disc in the presence of this
deformation.}
\end{quote}
}

\bigskip\bigskip


\end{titlepage}
\maketitle\vfill \eject

\tableofcontents

\sectiono{Introduction}\label{sec:Introd}
The study of tachyonic instabilities in configurations of D-branes in string theory
is an important one\cite{Sen:2004nf}. While time-dependent processes in string
theory are difficult to analyze in general, the process of tachyon rolling down the
potential is one in which certain aspects of dynamics have been found to be
tractable\cite{Moeller:2002vx,Hellerman:2008wp,Joukovskaya:2008zv,%
Barnaby:2008pt,Beaujean:2009rb,Song:2010hc}.
In particular, the pioneering studies in Ref.\cite{Moeller:2002vx} found solutions
of the equations of motion of cubic, open string field theory that move away from
the unstable vacuum, but also go past the stable vacuum and exhibit wild oscillations
at late times. This is not unexpected in the absence of any coupling to the closed
string modes, to which an unstable D-brane is expected to decay\cite{Sen:2004nf}.
A complete treatment of this problem would require one to use interacting open-closed
string field theory\cite{Zwiebach:1997fe}, the formulation of which is not well
understood as yet (see, however, Refs. \cite{Ohmori:2003je,Ishida:2012cw}).

An alternative, which avoids this complexity, is to put the open strings in the
presence of a linear dilaton background, which of course originates in the
closed string sector. This was suggested by the authors of
Ref.\cite{Hellerman:2008wp}, who use light-cone coordinates $x^\pm$, and
consider the dilation profile linear in $x^-$. The underlying conformal field
theory (CFT) is well understood: the dilaton, being linear along a null
direction, changes the (world-sheet) conformal dimension of the vertex operators,
but does not alter the matter contribution to the central charge.
Ref.\cite{Hellerman:2008wp} studied the homogeneous decay\footnote{In
the terminology adopted in this paper, `homogeneous' refers to processes
dependent on light-cone time ($x^+$) only. We will also use $x^+\equiv\tau$
to simplify notation.} of the tachyon as a function of  light-cone time $x^+$,
solved the equation of motion for the tachyon (zero-level truncation) and
extended this to the equations of motion of the full open string field theory.

The case of inhomogeneous decay in this framework was considered in
Ref.\cite{Ghoshal:2011rs}. At zero level truncation, the equation of motion of the
tachyon (as a function of $\tau$ and one other coordinate $y$ along the brane)
was found to have a close resemblance to the ubiquitous reaction-diffusion
equation pioneered in Refs.\cite{Luther,Fisher, KPP}. Specifically, the non-linear
reaction term of the `Fisher equation for the tachyon on a decaying brane',
Eq.\refb{pheom-dil}, involves a time delay and spatial averaging with a
Gaussian kernel, hence it is non-local\footnote{Non-locality in reaction-diffusion
systems has been considered in subsequent literature, mainly in Mathematical
Biology. However, the specific form that appears in the tachyon equation of motion
is distinct to our knowledge.} as well. Like its ancestor, the Fisher equation for the
tachyon has a {\em travelling front} solution that separates the brane from the
(closed string) vacuum and moves with a constant speed retaining its shape.
This solution was found using a singular perturbation analysis.

In this paper, we discuss the extension of this travelling front to the equations of
motion of open string field theory (which takes the effect of the higher stringy
modes into account). Specifically, we start with the deformation corresponding
to the front solution in Ref.\cite{Ghoshal:2011rs}---this is a marginal perturbation
of the D-brane CFT. In fact, there is a continuous family of marginal operators,
however, we shall see that only one of these allows for a solution to the equations
at second order. It is the same one for which the front propagates with the
minimum speed. Thus, in both situations this operator plays a special role. It
seems likely that it is exactly marginal. We call the corresponding perturbation
the {\em Fisher deformation}.

In the following, we start with a brief review of some relevant results from the
literature. In Sec.\ref{sec:sfteom}, we extend the marginal Fisher deformation to
the equations of motion of string field theory to the second order in perturbation
parameter. We comment on the gauge condition and the complications in
extending to higher order terms. Sec.\ref{sec:DiscPF} deals with the one-point
function of closed string vertex operators in the presence of Fisher deformation
of the boundary CFT. We end with a summary and some comments.

\sectiono{Travelling front to the tachyon equation: a review}\label{sec:Rev}
Let us, for definiteness, consider the CFT corresponding to an unstable D$p$-brane
of the bosonic string theory. The string field $\left|\Psi\right\rangle$ is a vector
in the Hilbert space of the matter-ghost (boundary) CFT, and may be expanded as
\begin{equation}
\left|\Psi\right\rangle = \phi(X) c \left| 0\right\rangle + \cdots =
\int \frac{d^{p+1}k}{(2\pi)^{p+1}} \phi(k) e^{ik.X} c_1 \left| 0\right\rangle + \cdots ,
\label{PsiExpanded}
\end{equation}
where $\phi$ is the tachyon and the dots stand for the higher stringy modes.
The Chern-Simons type action of the cubic open string field theory
\begin{equation}
S = \frac{1}{g^2}\left( \frac{1}{2} \left\langle\Psi\right|Q_B\left|\Psi\right\rangle
+ \frac{1}{3} \langle\Psi | \Psi\star \Psi\rangle\right),
\label{osftAction}
\end{equation}
is defined in terms of the $\star$-product, the BRST operator $Q_B$ and the
inner product of the matter-ghost CFT. If we retain only the tachyon field $\phi$
(level truncation to zeroth order) and further restrict to spatially homogeneous
decay, {\em i.e.}, $\phi$ depends only on time $t$, the equation of motion has 
solutions that start at the maximum of the potential (at $\phi_U=0$) towards the 
(local) minimum (at $\phi= K^{-3} \simeq 0.456$), but overshoot and exhibit
(non-linear) oscillations around the minimum. At late times, as a result of the
non-local non-linear interactions, these behave wildly\cite{Moeller:2002vx}.
A solution that interpolates between the D-brane and the (closed string)
vacuum has not been found. Physically this is not unexpected, as the energy
of the D-brane cannot dissipate into the closed string modes in a theory of
open strings alone.

Let us now consider a linear dilaton background
${\cal D}(x)= -D^+x^- \equiv -b x^-$ (where $x^\pm=(t \pm x)/\sqrt{2}$ and $b$
is a constant) following Ref.\cite{Hellerman:2008wp}. The linear dilaton CFT
is solvable. Moreover, since the dependence of the dilaton is along a null
direction, the central charge of the matter CFT remains the same, Only the
(world-sheet) conformal dimension of the tachyon vertex operators $e^{ik.X}$
change from $k^2$ to $k^2 + ibk^-$. Consequently, the equation of motion for
the tachyon gets modified to (we have used $\alpha'=1$):
\begin{equation}
\left(b\,{\partial\over\partial\tau} - \nabla^2_\perp\right)
\phi(\tau,\mbox{\bf x}_\perp) =
\phi(\tau,\mbox{\bf x}_\perp) - K^3 e^{-2\alpha b\partial_\tau +
\alpha\nabla^2_\perp} \left[e^{\alpha\nabla^2_\perp}
\phi(\tau,\mbox{\bf x}_\perp)\right]^2,
\label{pheom-dil}
\end{equation}
where $K = 3\sqrt{3}/4$, $\alpha = \ln K$ and $\mbox{\bf x}_\perp$ denotes the
coordinates along the D-brane that are transverse to the light-cone coordinates. This
has been referred to as the `Fisher equation for the tachyon on a decaying brane'
in \cite{Ghoshal:2011rs}.

\begin{figure}[ht]
\begin{center}
\includegraphics[scale=0.8]{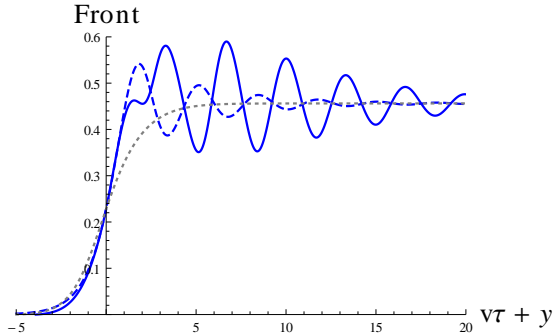}
\end{center}
\caption{{\small
The interpolating solution $\phi(\tau)$ of the the homogeneous tachyon equation,
which also solves the lowest order inhomogeneous equation, is the blue dashed
curve. With the first corrections it is the solid blue curve. The solution of the 
ordinary Fisher equation (gray dotted) is shown for comparison. (The argument is
only $\tau$ for the homogeneous case.)
}}
\label{fig:TrFront}
\end{figure}

In homogeneous decay, $\phi=\phi(\tau)$ depends only on (light-cone) time, therefore,
the equation simplifies\cite{Hellerman:2008wp,Barnaby:2008pt,Beaujean:2009rb,%
Song:2010hc} to that of a delayed growth model\cite{MurrayMB}. The usual growth
model (logistic equation) of population dynamics has a simple interpolating solution,
but the delay leads to oscillations around the stable fixed point at $\phi_S$. The solution
interpolating between these fixed points (see Fig.\ref{fig:TrFront}) was found in
\cite{Hellerman:2008wp} (see also \cite{Ghoshal:2011rs}).

Now consider the case of inhomogeneous decay governed by the Fisher equation
\refb{pheom-dil}. Let us, for simplicity, take the tachyon to depend only on one
direction along the brane, and denote this by $y$ (it is transverse to the light-cone
coordinates $x^\pm$). There is a {\em travelling front} solution that moves, say, from
right to left, so that at any instant of time the region to the right of the front moves
towards the stable fixed point. If we linearize the equation around the maximum
$\phi_U=0$, and put in the ansatz $\phi_v \sim \exp\left(k(y + v(k)\tau\right)$, we
find the dispersion relation
\begin{equation}
v(k) = {1\over b}\left(k + {1\over k}\right).
\label{lin_disp}
\end{equation}
The wavenumber $k$ is real for $v(k)\ge v_{\mathrm{min}} = 2/b$, therefore, any of
these solves the linearized equation. For a large class of nonlinear interactions, the
travelling front of the usual Fisher-type equation (without delay or nonlocality), has
been proven to select the front solution corresponding to
$v_{\mathrm{min}}$\cite{MurrayMB,LDNath}. Ref.\cite{Ghoshal:2011rs} made the
plausible assumption\footnote{The additional elements of delay and nonlocality
do not affect the linearized analysis around the maximum of the tachyon potential.
The `leading edge' of the wave is determined by the `mass' of the tachyon and the
parameter $b$, as in the standard case.} that this feature is also true of
Eq.\refb{pheom-dil}. Then a travelling front solution to this equation in the form of
$\phi(\tau,y) = \Phi(\eta = y + v\tau)$, was found by a straightforward adaptation of
singular perturbation analysis\cite{MurrayMB,LDNath} in terms of the parameter
$\varepsilon\equiv 1/v^2b^2 \le 0.25$. At the leading order, {\em i.e.}, ${\cal O}(1)$
in $\varepsilon$, the front is just as in Fig.\ref{fig:TrFront}---higher order corrections,
can be found systematically following Ref.\cite{Ghoshal:2011rs}.

\sectiono{Travelling front in OSFT}\label{sec:sfteom}
The approach of Ref.\cite{Ghoshal:2011rs} outlined above, however, gives us a
solution to the tachyon equation of motion \refb{pheom-dil}, which is an approximate
solution, being a level-zero truncation of the equation of motion of string field theory
\[
Q_B\left|\Psi\right\rangle + \left| \Psi\star \Psi\right\rangle = 0.
\]
We would now like to address the question of finding a solution to the above,
starting with the tachyon vertex operator
\begin{equation}
\phi_k = :\exp\left(k(Y + v(k)X^+)\right):
\label{fisher_def}
\end{equation}
with $v(k)$ as in Eq.\refb{lin_disp}.
Clearly, $\Psi = c\phi_k$ solves the linearized equation of motion
$Q_B|\Psi\rangle = 0$. This is equivalent to the statement that $\phi_k$ is a
marginal deformation of the underlying (boundary) CFT. Indeed, the dimension of
$\phi_k$ (in the linear dilaton background) is
\begin{equation}
h(\phi_k) = k_\mu k^\mu + i k_\mu D^\mu = - k^2 + i (-ikv(k)) b = 1
\label{dimFish}
\end{equation}
for all values of $k$, thanks to the dispersion relation \refb{lin_disp}. (Let us note
parenthetically that this is reminiscent of `Liouville dressing' of matter vertex
operators in non-critical string theory.) At this stage $\phi_k$ with any $k$ seems
to be a good marginal deformation. However, we shall see later that the value
$k=1$, for which $v(k=1)=v_{\mathrm{min}}$, and therefore, plays a special role
in the travelling front solution to the Fisher equation of the tachyon, turns out to be
special as a marginal deformation. We shall refer to $\phi_{k=1}$ as the
{\em Fisher deformation}.

Let us note that the case of homogeneous decay is recovered with $k=0$ and
$kv(k) = 1/{b}$. The marginal deformation corresponding to the tachyon vertex
operator $e^{X^+/b}$ in this case is special in that its OPE with itself vanishes
identically. This considerably simplifies the problem of extending this marginal
deformation to a solution of the SFT equations of motion\cite{Hellerman:2008wp}.

We shall closely follow the method (and the notations) of Ref.\cite{Kiermaier:2007ba}
(see also \cite{Schnabl:2005,Okawa:2006,OkawaRasZwie:2006}) in order to extend
the marginal Fisher deformation \refb{fisher_def} to a solution of the SFT equation of
motion. To this end, let us define
\[
\Psi_\lambda = \sum_{n=1}^\infty \lambda^n\, \Psi^{(n)} = \lambda\,c\,\phi_k +
\lambda^2\, \Psi^{(2)} + \cdots,
\]
which allows one to determine $\Psi^{(n)}$ iteratively from
\begin{equation}
Q_B\Psi^{(n)} = - \displaystyle\sum_{m=1}^{n-1} \Psi^{(m)}\star\Psi^{(n-m)}
\label{rec_soln_sft}
\end{equation}
and construct the solution as a power series in $\lambda$. However, this
involves inverting $Q_B$. It is by now well established that the solution is
best attempted in the sliver frame in the Schnabl gauge. Formally, it is given by
\begin{equation}
\left\langle\omega,\Psi^{(n)}\right\rangle = \prod_{m=1}^{n-1} \int_0^1\!\! dt_m\;
\left\langle f\circ\omega(0)\, c\phi_k(1)\, {\cal B}c\phi_k(1+t_1) \cdots
{\cal B}c\phi_k\left(1+\sum_m t_m\right) \right\rangle
\label{SlivrSolLevln}
\end{equation}
evaluated on the surface ${\cal W}_{1+\sum t_m}$, a
wedge state in the sliver frame. In the above, $\omega$ is a generic state in the
Hilbert space and $f(z) = \frac{2}{\pi}\tan^{-1}(z)$ is the conformal map from the
upper half plane to the sliver.

This expression is formal because of possible singularities that can arise when
two operator insertions collide. In the case of the homogeneous decay considered
in \cite{Hellerman:2008wp}, the null field $X^+$ does not have a contraction with
itself, as a result of which there is no singularity when the marginal operators
collide, and the formal solution above is well defined. In the case of Fisher
deformation however, the OPEs are singular and regularization is needed to
make sense of \refb{SlivrSolLevln}.

\subsectiono{Solution at second order}
The first correction $\Psi^{(2)}$ obtained from \refb{SlivrSolLevln} is ill-defined due
to the singularity from $t\to 0$. Let us regularize this as
\begin{equation}
\left\langle\omega,\Psi^{(2)}_{\mathrm{reg}}\right\rangle =
\lim_{\epsilon\to 0} \int_{2\epsilon}^1\!\! dt\; \left\langle f\circ\omega(0)\,
c\phi_k(1)\, {\cal B}c\phi_k(1+t) \right\rangle_{{\cal W}_{1+t}}.
\label{RegLevl2}
\end{equation}
The above has a finite part and a divergent part, so we write
$\Psi^{(2)}_{\mathrm{reg}} = \Psi^{(2)}_{\mathrm{fin}} + \Psi^{(2)}_{\mathrm{div}}$.
The finite part is as in Ref.\cite{Kiermaier:2007ba}, hence can be dealt with as in
there. The divergent part comes from the region $\epsilon\to 0$, where we use
the OPE of the vertex operators to write
\begin{equation}
\begin{split}
\lim_{\epsilon\to 0}\; c\phi_k(1) & {\cal B}c\phi_k(1+2\epsilon) \; = \\
& \left[\frac{1}{(2\epsilon)^{2k^2}} :\!c\phi^2_k\!: + \frac{1}{(2\epsilon)^{2k^2-1}}
\left(\frac{1}{2}:\!c\partial\phi_k^2\!:
- :\!c \partial c \phi^2_k {\cal B}\!: + :\!c \phi^2_k {\cal L}\!: \right)(1)+\cdots \right],
\end{split}
\label{opeLevl2}
\end{equation}
where $:\!\phi_k^2\!: \, =\, :\!e^{2k\left(Y - v(k)X^+\right)}\!:$ has dimension $2-2k^2$.
(We note, in passing, a curiosity of these deformations: Even though
the perturbations are marginal, their behaviour in the OPE is like those of relevant
ones---an exception being the Fisher deformation, which behaves like a truly
marginal operator.) From the above, we find
\begin{equation}
\Psi^{(2)}_{\mathrm{div}} (\epsilon)  = \frac{c:\!\phi_k^2\!:}{(2k^2-1)
(2\epsilon)^{2k^2-1}} + \int_{2\epsilon} \frac{dt}{t^{2k^2-1}}
\left(\frac{1}{2}c:\!\partial\phi^2_k\!:  -\, c \partial c {\cal B} :\!\phi^2_k\!:
+\, c :\!\phi^2_k\!:{\cal L} \right) + \cdots
\label{Psi2div}
\end{equation}
Notice that there is no divergence for $k^2 < {1}/{2}$. For $k^2=1/2$, the first term
is logarithmically divergent, therefore, the renormalization method will fail. Both the
terms are divergent for  $\frac{1}{2} < k^2 < 1$, while for $k^2 = 1$, namely for the
Fisher deformation, there is a logarithmic divergence in the second term. Finally, for
$k^2 > 1$, there will be additional divergent terms.

We shall return to the first term, but first we want to check if the second term is
BRST-exact. Indeed it is easy to check that
\begin{eqnarray}
\frac{1}{2}c:\!\partial\phi^2_k\!:  -\, c \partial c {\cal B} :\!\phi^2_k\!: +\,
c :\!\phi^2_k\!: {\cal L}
&=& Q_B \left(\frac{1}{2}:\!\phi^2_k\!: - \, c{\cal B}:\!\phi^2_k\!: \right)
\label{logdivterm}\\
&{}&\: +\, (2k^2-2) \left(\frac{Q_B\left(c:\!\phi^2_k\!:\right){\cal B}} {2k^2-1} +
\frac{1}{2} \partial c :\!\phi^2_k\!: \right) .
\nonumber
\end{eqnarray}
{\em i.e.}, a part of it is independent of $k$ and BRST-exact, however, there is in
general also a $k$-dependent piece which spoils BRST-exactness. The only
exception is the case of Fisher deformation $\phi_{k=1}$, for which the second term
vanishes, therefore, the operator is BRST-exact. At this point we can use an ambiguity 
in $\Psi^{(2)}$. As one can see from Eq.\refb{rec_soln_sft} it is defined only upto a 
BRST-closed term. Therefore, we are free to  add to it $\Psi^{(2)}_{\mathrm{exact}}$, 
a BRST-exact term defined as
\begin{equation}
\Psi^{(2)}_{\mathrm{exact}} =  - \int_{2\epsilon} \frac{dt}{t^{2k^2-1}} Q_B
\left(\frac{1}{2}:\!\phi^2_k\!: - \, c{\cal B}:\!\phi^2_k\!: \right)
\label{Psi2exact}
\end{equation}
to remove the second divergent term in Eq.\refb{Psi2div}.

As for the first divergent term, following \cite{Kiermaier:2007ba}, we notice that
the regularized expression \refb{RegLevl2} does not satisfy the SFT equation of
motion due to the presence of the surface term
\begin{equation}
\begin{split}
\left< f \circ \omega(0)\right. & \left.c\! :\!\phi_k\!:\!(1)\, c\!:\!\phi_k\!:\!(1+2\epsilon)
\right>_{W_{1+2\epsilon}} \label{divSurfc}\\
& = \left<f \circ \omega(0)\; Q_B\left( \frac{1}{(2\epsilon)^{2k^2-1}}
  \frac{c:\!\phi^2_k\!: (1+\epsilon) }{2k^2-1} + \frac{{\cal B} c \partial c
  :\!\phi^2_k\!: (1+\epsilon) }{(2\epsilon)^{2k^2-2}} +
  \cdots\right)\right>_{{\cal W}_{1+2\epsilon}}\\
& \equiv -\, \left<f \circ \omega(0)\; Q_B\Psi^{(2)}_{\mathrm{CT}}(\epsilon)\right>.
\end{split}
\end{equation}
Consequently,
\[
\Psi^{(2)}_{\mathrm{CT}}(\epsilon) = -\,\frac{1}{(2k^2-1)(2\epsilon)^{2k^2-1}}\,
{c\!:\!\phi^2_k\!:(1+\epsilon)}
\]
(defined upto terms which are regular for $\frac{1}{2} < k^2 \leq 1$) may be used as a
counter-term in defining $\Psi^{(2)}$. This counter-term exactly cancels the the divergent
first term in Eq.\refb{Psi2div}. Hence, the renormalized string field
\[
\Psi^{(2)}_{\mathrm{ren}} = \lim_{\epsilon\to 0}\,\left(\Psi^{(2)}_{\mathrm{reg}} +
\Psi^{(2)}_{\mathrm{exact}} + \Psi^{(2)}_{\mathrm{CT}}\right)
\]
is finite and satisfies the equation of motion to ${\cal O}(\lambda^2)$.

Although we are able to follow the steps in Ref.\cite{Kiermaier:2007ba} closely, the
situation here is different from the one considered there. The general formalism
assumes that the OPE of the marginal deformation is either regular, or has a double
pole, but not a simple pole term. In contrast, the OPE of the Fisher deformation does
have a simple pole term. Thankfully, however, the corresponding operator turns out
to be BRST-exact (see Eq.\refb{logdivterm} above and the remarks in
\cite{Kiermaier:2007ba}), hence the ideas developed there also work in this case.

\subsectiono{Gauge condition}
The solutions outlined in Ref. \cite{Kiermaier:2007ba} break the (Schnabl) gauge
condition. This is also the case for the solution seeded by the Fisher deformation.
Specifically, the counter-term $\Psi^{(2)}_{\mathrm{CT}}$ breaks the gauge condition.

First, note that $\Psi^{(2)}_{\mathrm{reg}}$ and $\Psi^{(2)}_{\mathrm{exact}}$ satisfy
the gauge condition:
$B \Psi^{(2)}_{\mathrm{reg}} =0$
and  $B \Psi^{(2)}_{\mathrm{exact}} = 0$,
where $B$ is the zero mode of the anti-ghost in the sliver frame. Recall that
$c = \frac{2}{\pi}c_1$, and $B^+ \equiv B+B^\flat$, where $B^\flat$ is BPZ conjugate
of $B$. Using the short-hand $:\!\! \phi^{2}_{k=1}\!\!:\, \equiv \phi_F^{2}$, the
counter-term $\Psi^{(2)}_{\mathrm{CT}}$, can be written as
\begin{equation}
\Psi^{(2)}_{\mathrm{CT}} = - \frac{1}{\pi \epsilon} e^{\epsilon L^+}
c_1 \phi_F^2 \left| 0 \right>
=  -\frac{1}{\pi \epsilon} c_1 \phi_F^2 \left| 0 \right>  +
\frac{1}{\pi} L^+ c_1 \phi_F^2 \left| 0 \right>  +  {\cal O}(\epsilon),
\end{equation}
where $L$ and $L^\flat$ are the zero mode of the energy-momentum tensor and its
BPZ conjugate in the sliver frame, and $L^+ \equiv L + L^\flat$. It is not difficult to
check that $ L^+ \phi_F^2 = 0$. This is due to fact that the conformal dimension
of $\phi_F^2$ is zero. Applying $B$ to the counter-term
\begin{equation}
B \Psi^{(2)}_{\mathrm{CT}}
=\frac{1}{\pi} B^+ c_1 \phi_F^2 \left| 0 \right> \ne 0,
\end{equation}
we see that the gauge condition is violated by the counter-term. One can again take
advantage of the ambiguity in the solution of \refb{rec_soln_sft} and add a BRST-closed
state $\Xi$ and try to restore the gauge condition. If such a state exists, then
$B(L^+ c_1 \phi_F^2 \left| 0 \right>  + \Xi)=0$. When applied with $Q_B$, we find
\begin{equation}
- L\, \Xi =  L L^+ c_1 \phi_F^2 \left| 0 \right>  - B Q_B L^+ c_1 \phi_F^2 \left| 0 \right>
=  - B Q_B L^+ c_1 \phi_F^2 \left| 0 \right>,
\label{kernelL}
\end{equation}
where in addition to the standard relations $\{Q_B,B\}=L$ and $[L,L^+]=L^+$, we have
used $L c_1 \phi_F^2 \left| 0 \right> = - c_1 \phi_F^2 \left| 0 \right>$. Once again, the
right hand side is in the kernel of $L$, since $ L L^+ c_1 \phi_F^2 \left| 0 \right> = 0$.
In fact, the only difference from Ref. \cite{Kiermaier:2007ba} is in the appearance of
$\phi_F^2$ in place of the identity operator. However, this does not change anything
because it is also a dimension-zero operator.

The formal solution $\Psi^{(n)} = \frac{B}{L} \Phi^{(n)}$ in the Schnabl gauge is
well-defined when $\Phi^{(n)}$ does not have an overlap with the kernel of $L$.
Moreover, $\Phi^{(2)}$ must be even under ghost-twist because $\Psi^{(1)}$ is even.
There is a ghost-number two, even, BRST-exact term of dimension zero
\begin{eqnarray}
L^+ c_1 c_0 \phi_F^2 \left| 0 \right>  = Q_B L^+ c_1 \phi_F^2 \left| 0 \right>
\end{eqnarray}
in the subspace of states for $\Phi^{(2)}$. This term is not annihilated by $B$, therefore,
breaks the gauge condition.

\subsectiono{Comments on higher order correction}
The computations for $\Psi^{(3)}$ at the next order (and beyond) get rapidly very complicated,
even for the Fisher deformation. This is due to the non-vanishing simple pole term in its OPE.
As a result, the most singular term in the OPE of $\phi_F$ and $\phi_F^2$ has a fourth order
pole! More generally,
\begin{equation}
:\!\phi^2_k\!:\!(z)\, :\!\phi_k\!:\!(w) = \frac{1}{(z-w)^{4k^2}}\, {:\!\phi_k^3\!:} (w) +
\frac{2}{3(z-w)^{4k^2-1}}\, {\partial_w\!:\!\phi_k^3\!:}(w) + \cdots.
\end{equation}
Consequently, while renormalizing the formal solution, we encounter
\begin{equation}
:\!c\phi^2_k\!:\!(1+\epsilon)\, \:\!c\phi_k\!:\!(1+4\epsilon) = \frac{1}{(3 \epsilon)^{4k^2-1}}\,
{:\!c \partial c \phi_k^3\!:} + \cdots.
\end{equation}
The corresponding operator for the Fisher deformation is, however, BRST-exact:
$c \partial c \phi_F^3 = \frac{1}{4}Q_B(c \phi_F^3)$.
A counting of dimensions show that the dimensions of the operators accompanying all the
singular terms are non-zero (negative) integers. In particular, the OPEs of $\phi_F$ and
$\phi_F^2$ do not produce another marginal operator. This is also the case for the OPEs
of $\phi_F^n \sim :\! e^{nk(Y + v(k)X^+)}\!:$ (the dimension of which is $-n(n-2)$) at higher 
order, from the same dimension counting argument. Thus, it seems possible that all the 
singular terms are BRST-exact for the Fisher deformation, hence a renormalized solution 
may be found. However, we shall not attempt to prove this here.

\sectiono{Disc partition function}\label{sec:DiscPF}
We shall now turn to the computation of the partition function on the disc with an
insertion of a closed string vertex operator in the interior, in the presence of the
Fisher deformation of the boundary CFT. This computation has been done many
times in the past, {\em e.g.}, in Refs.\cite{Larsen:2002wc,Lambert:2003zr} for
timelike tachyon (with or without additional spatial dependence) and in
Refs.\cite{Hellerman:2008wp,Song:2010hc} for lightlike tachyon in a linear
dilaton background. One can make use of the result  in different ways: the
partition function on the disc is closely related to the (space-time) action in the
boundary string field theory approach\cite{Witten:1992qy,Witten:1992cr}---the
two actually coincide on-shell. Moreover, one can get the energy-momentum tensor
\cite{Sen:2002nu} by choosing the graviton vertex operator in the interior.

Let us first calculate the one-point function of a closed string tachyon vertex operator
with the momentum $p_\mu = \left(p_+,p_-,0\right)$
\[
V_{\mathrm{closed}}(X) \equiv V(p_+,p_-) = e^{p_+X^+ + p_-X^-}
\]
on the disc $\left< V(p_+, p_- )\right>_{\mathrm{Disc}}$ in the background
of the lightlike linear dilaton, in the presence of the perturbation $\int\phi_k(X)$ to
the boundary CFT. We separate the zero modes $x^\mu$, and fix these with the
using $\delta\left(x^\mu - \int\frac{d\theta}{2\pi}X^\mu\right)$. Essentially we do
the computation in \cite{Hellerman:2008wp,Song:2010hc} for the deformation
$\phi_k=:\!e^{k(Y+v(k)X^+)}\!:$, and also use the normal ordering described there.
The result, from conformal perturbation theory, is
\begin{eqnarray}
&{}& \left< V(p_+, p_- \right>_D \nonumber\\
&=&  \int d x^+ \sum^{\infty}_{n =0} \frac{(- \lambda)^n}{n !} \left(\prod^n_{i=1}
\int^{2\pi}_{0}\!\frac{d\theta_i}{2 \pi}\right)  \left<V(p_+,p_-)
\prod^{n}_{i=1}  :\!e^{k(Y+v(k)X^{+})} \!: \right> \label{pert1ptfn} \\
&=& \int d x^+ e^{p_+ x^+ + p_- x^-}
\sum^{\infty}_{n =0} \frac{(- \lambda)^n}{n !}  e^{nk(v(k) x^{+} + y)}
\left(\prod^n_{i=1}\int^{2\pi}_{0}\! \frac{d\theta_i}{2 \pi}\right)
\prod_{1\leq i < j \leq n} \left| e^{i \theta_i} - e^{i \theta_j}\right|^{-2k^2} .
\nonumber
\end{eqnarray}
Notice that the power of the separation between the points of the boundary of
the disc is negative. As a result, the integrand diverges whenever two (or more)
boundary operators coincide. This is unlike in Refs.\cite{Hellerman:2008wp,%
Song:2010hc} and Ref.\cite{Larsen:2002wc}. In the case of the former, the integrand
is trivial, because the lightlike tachyon deformation does not have a non-zero
self-contraction, while, in the latter case of timelike tachyon, the power is positive.

The divergence in expression \refb{pert1ptfn} needs to be regulated. One way
would be to use point splitting, which would break conformal invariance. This
requires one to proceed carefully. Instead, we recall that integrals of this type
appear in the theory of random matrices (RMT). Indeed the authors of
Ref.\cite{Larsen:2002wc} used this observation to evaluate their integrals. In the
context of RMT, these are known as Dyson's integrals, the values of which were
conjectured by Dyson\cite{Dyson} to be
\begin{equation}
{\cal D}_n(\beta) = \left(\prod^n_{i=1}\int^{2\pi}_{0}\! \frac{d\theta_i}{2 \pi}\right)
\prod_{1\leq i < j \leq n} \left| e^{i \theta_i} - e^{i \theta_j}\right|^{2\beta} =
\frac{\Gamma(1 + n\beta)}{\left(\Gamma(1 + \beta)\right)^n}
\label{DysonInt}
\end{equation}
and were proved in Refs.\cite{DysonProof}. The function ${\cal D}_n(\beta)$ is 
a meromorphic function of $\beta$. For the tachyon deformation considered here, 
we have ${\cal D}_n(-k^2) = \Gamma(1-nk^2)/\left(\Gamma(1-k^2)\right)^n$, 
leading to the series
\begin{equation}
\sum^{\infty}_{n =0} \frac{(- \lambda)^n}{n !}\,  \frac{\Gamma(1-nk^2)}{\left(
\Gamma(1-k^2)\right)^n}\, e^{nk(y + v(k) x^{+})}.
\label{DysonSeries}
\end{equation}
For the homogeneous case, $k^2=0$, hence this is just the exponential series and
known results \cite{Song:2010hc} are recovered. For a general inhomogeneous
decay, this is at best an asymptotic series: the numerator diverges at order $n=q$
for $k^2={p}/{q}$.

In the case of Fisher deformation ($k^2=1$), both the numerator and the
denominator are singular for $n\ge 2$, however, due to the presence of multiple
factors of $\Gamma(0)$ in the denominator, the integrals actually vanish for
$n\ge 2$! Taking this at face value, the only contributions are from the first two
terms. This gives us $1 - \lambda \exp\left({y + \frac{2}{b}x^+}\right)$, which goes
to 1 as the argument of the exponential goes to $-\infty$ (D-brane background)
but diverges (to $-\infty$) as it goes to $\infty$ (closed string vacuum). Clearly,
this is not the expected behaviour.

\begin{figure}[ht]
\begin{center}
\includegraphics[scale=0.8]{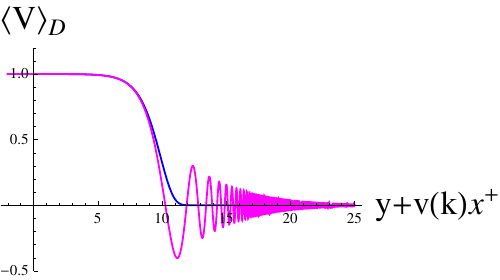}
\hspace*{18pt}
\includegraphics[scale=0.7]{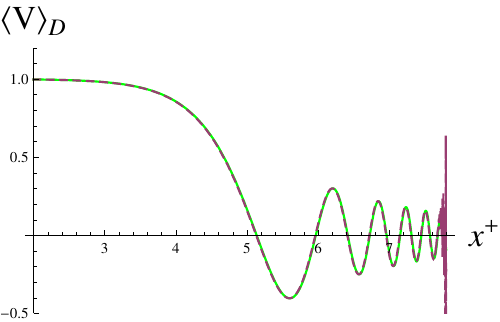}
\end{center}
\caption{{\small
On the left: Oscillatory decay of $\left<V\right>_D$ as given by the Bessel function
in Eq.\refb{FishDef1ptRen2}. The corresponding result for homogeneous decay is
shown in solid blue.
On the right: A comparison of the perturbative series with the expression
\refb{FishDef1ptRen2}. Notice that the series starts to behave badly for large
values of the argument.}}
\label{fig:TachDisc}
\end{figure}

Let us, instead, set $k^2 = 1 + \varepsilon$, so that
${\cal D}_n(-k^2 = -1-\varepsilon) = \frac{(\varepsilon)^{n-1}}{n!}$,
using which we find
\begin{equation}
\left<V\right>_D \equiv \left< V(0, 0 \right>_D \sim \frac{1}{\varepsilon}\,
\sum_{n=0}^\infty\frac{1}{(n!)^2}\,\left(-\varepsilon\lambda e^{y + \frac{2}{b}x^+}\right)^n
=  \frac{1}{\varepsilon}\, J_0\left(2\sqrt{\varepsilon\lambda}\,
e^{\left(y + \frac{2}{b}x^+\right)/2}\right),
\label{FishDef1ptRen}
\end{equation}
where $J_0(z)$ is the Bessel function. We get a finite answer if we renormalize the
coupling $\lambda_R = \varepsilon\lambda$. The additional factor of $\varepsilon$
may be absorbed in the coupling of the closed string vertex operator $V$. In terms
of the renormalized couplings
\begin{equation}
\left< \tilde{V}\right>_D \sim  \, J_0\left(2\sqrt{\tilde\lambda}\,
\exp{\frac{1}{2}\left(y + \frac{2}{b}x^+\right)}\right),
\label{FishDef1ptRen2}
\end{equation}
which goes to one as the argument of the exponential goes to $-\infty$ (D-brane
background) and settles to zero (after some oscillations---see Fig.\ref{fig:TachDisc})
as it goes to $\infty$ (closed string vacuum).

Some comments are in order. First, we do not understand why the disc partition function
oscillates around the closed string vacuum. As such these do not seem to be related
to the oscillations in Fig.\ref{fig:TrFront}. Those are due to the time delay, and present
also for the homogeneous decay. Secondly, while the renormalization gives a sensible
result, it will be good to have a better understanding of its implications. Finally, if we 
approach $k^2=1$ from below, by parametrizing $k^2=1-\varepsilon$ instead, the 
Dyson's integrals alternate in sign: 
${\cal D}_n(-k^2 = -1+\varepsilon) =  \frac{(-\varepsilon)^{n-1}}{n!}$. 
The additional sign cancels the alternating signs in the perturbation series in $\lambda$ 
in \refb{DysonSeries}, and one ends up with the modified Bessel function of the first kind
$I_0\left(2\sqrt{\tilde\lambda}\,\exp{\frac{1}{2}\left(y + \frac{2}{b}x^+\right)}\right)$.
This does not oscillate, but diverges as the argument becomes large. It is worth noting,
however, that the integrands involved modulus-square of complex functions, thus are
manifestly positive, a feature that the parametrization we have used, preserves.

\begin{figure}[ht]
\begin{center}
\includegraphics[scale=0.30]{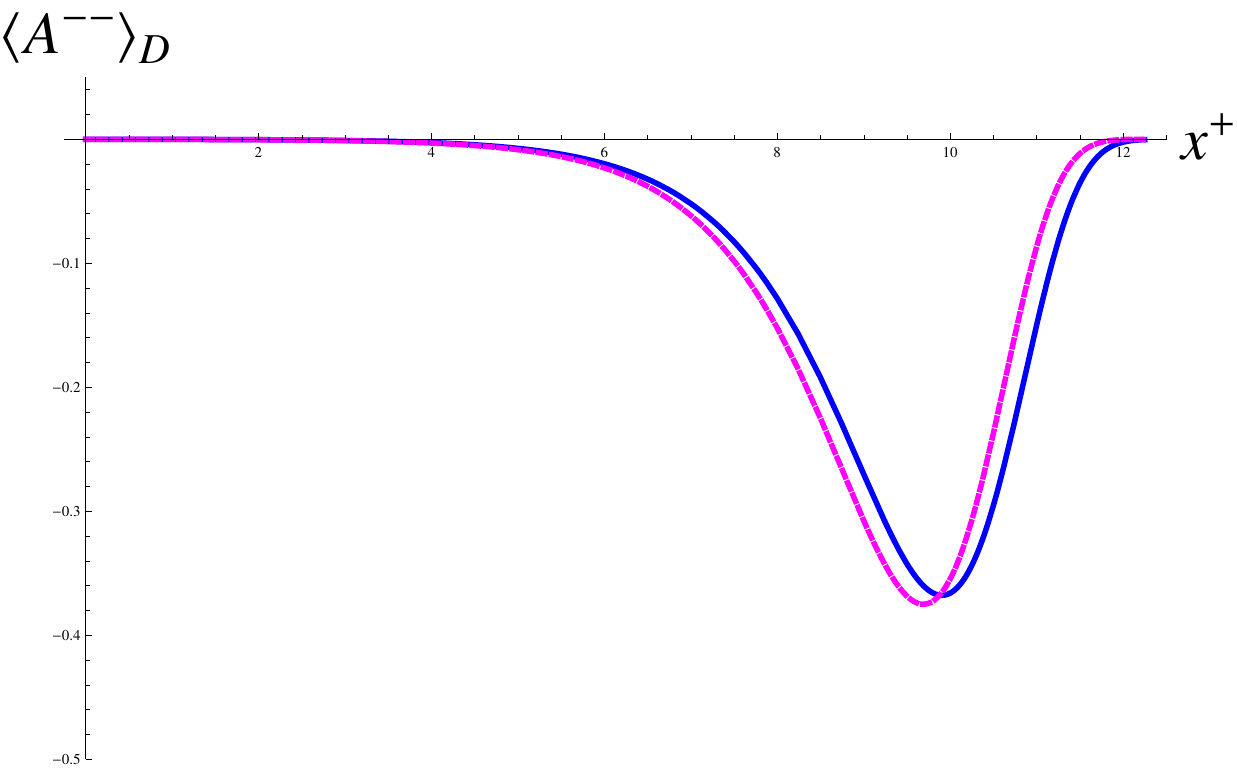}
\hspace*{18pt}
\includegraphics[scale=0.32]{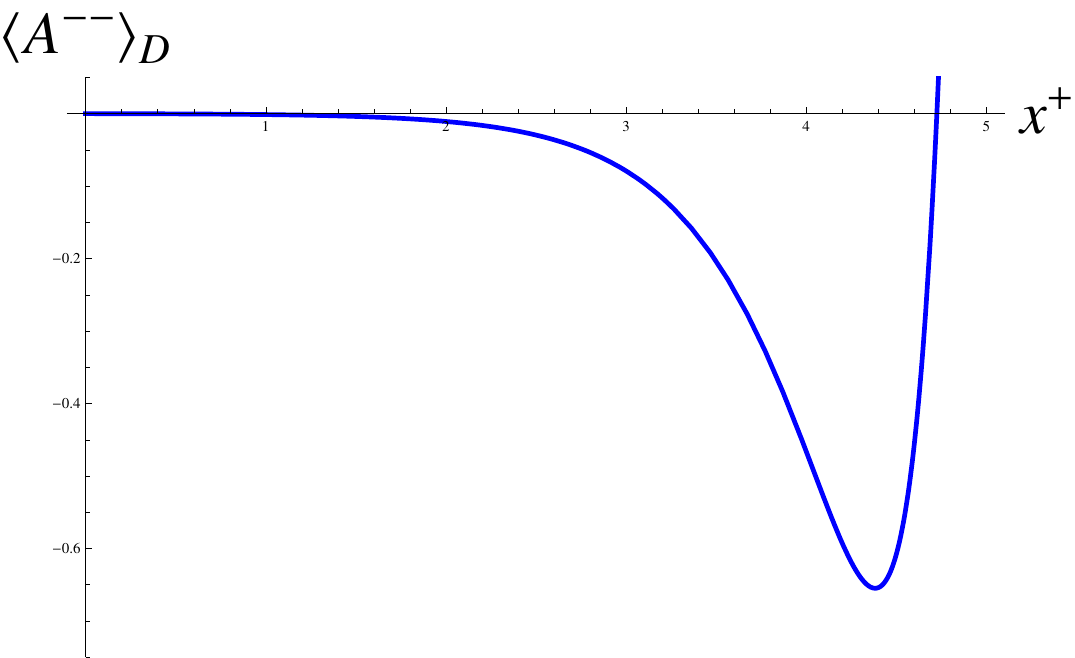}
\end{center}
\caption{{\small
The function ${\cal A^{--}}(-k^2)$ as a consequence of Eq.\ref{grav1pt}.
On the left: The solid (blue) curve is for the exact expression of ${\cal A^{--}}(0)$
and the dashed (magenta) curve is for the series with a small value of $k \sim 0.02$.
On the right: The plot of the series corresponding to the Fisher deformation, $k^2=1$.
It diverges for large values of the argument.}}
\label{fig:GravDisc}
\end{figure}

Next we choose the zero-momentum graviton for the closed string vertex operator, and
calculate the one-point function
\[
{\cal A}^{\mu\nu} = \left< : \!\partial X^\mu \bar{\partial} X^\nu\!: \right>_D
\]
on the disc. This requires evaluation of the integrals
\begin{equation}
{\cal A}_n(\beta) = \left(\prod^n_{i=1}\int^{2\pi}_{0}\! \frac{d\theta_i}{2 \pi}\right)
\prod_{1\leq i < j \leq n} \left| e^{i \theta_i} - e^{i \theta_j}\right|^{2\beta}
\sum_{\ell,m} e^{-i(\theta_\ell - \theta_m)}.
\label{modDysonInt}
\end{equation}
These integrals may be evaluated using the orthonormality of Jack 
polynomials\cite{Baker}.  (When separated into diagonal and off-diagonal parts, 
the former reduces to the Dyson integral.) The result is
\begin{eqnarray}
{\cal A}_n(\beta) &=& \left(\prod^n_{i=1}\int^{2\pi}_{0}\! \frac{d\theta_i}{2 \pi}\right)
\prod_{1\leq i < j \leq n} \left| e^{i \theta_i} - e^{i \theta_j}\right|^{2\beta}
\left( n + \sum_{\ell \neq m} e^{-i(\theta_\ell - \theta_m)} \right)  \nonumber \\
&=&  n \frac{\Gamma(1 + n\beta)}{\left(\Gamma(1 + \beta)\right)^n} -
\frac{\beta n(n-1)}{1+\beta(n-1)} \frac{\Gamma(1 + n\beta)}{\left(\Gamma(1 + \beta)\right)^n}
\nonumber \\
&=&  \frac{n}{1+\beta(n-1)} \frac{\Gamma(1 + n\beta)}{\left(\Gamma(1 + \beta)\right)^n}
\,=\, \frac{n}{1+\beta(n-1)} {\cal D}_n(\beta) .
\label{BakerInt}
\end{eqnarray}
This evidently agrees with the known result of Ref.\cite{Larsen:2002wc} for 
$\beta = 1$. We shall consider the above to be analytic in $\beta$ and continue to 
negative\footnote{The reciprocity relation for Gamma functions extends to the Beta 
functions and the Selberg integrals\cite{MimachiYoshida}. These are therefore 
well-defined for negative values of the parameters. Although we are not aware of 
such relations for the more general cases needed here, similar reciprocity relations 
are likely to be true.} values. In general, the associated series can be written as
\begin{equation}
{\cal A}(-k^2) = \sum_n\frac{(-\lambda)^n}{n!} e^{nk(y + v(k)x^+)}\, {\cal A}_n(-k^2).
\label{grav1pt}
\end{equation}
In the case of Fisher deformation, $\beta=-k^2=-1$, therefore, the additional pre-factor
in ${\cal A}_2$ is divergent. Writing $k^2 = 1 + \varepsilon$ as before, 
\begin{equation}
{\cal A}_2 
= 4~ \frac{\Gamma( -1 - 2 \epsilon))}{\left(\Gamma(-\epsilon)\right)^2} - 1 
=  2\varepsilon - 1.
\label{A2reg}
\end{equation}
We see that while the first term gives the expected form in terms of renormalized 
coupling $\lambda_R$ and renormalized closed string vertex operator, the second 
(constant) term will give a divergent contribution when written in terms of the renormalized 
quantities. This certainly is a cause of concern, and is perhaps due to the analytic 
continuation used. However, it is a problem for one term of the series and in the following, 
we shall omit this singular contribution and look at the rest of the series. Then, for the 
one-point function of the graviton vertex operators, we find that the non-vanishing 
components are
${\cal A}^{--} \sim k^2v^2\, {\cal A}(-k^2)$,
${\cal A}^{-y} \sim -k^2 v\, {\cal A}(-k^2)$ and
${\cal A}^{yy} \sim k^2\, {\cal A}(-k^2)$.
The behaviour of the function ${\cal A}^{--}$ given by the series expansion is shown in
Fig.\ref{fig:GravDisc}. For large values of the argument, the series is seen to diverge. 

\begin{figure}[ht]
\begin{center}
\includegraphics[scale=0.90]{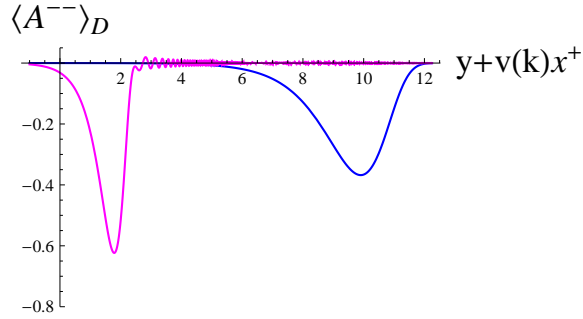}
\end{center}
\caption{{\small
The function ${\cal A^{--}}(-k^2)$ when the series in Eq.\ref{grav1pt} is re-expressed 
in terms of Bessel functions. The solid magenta curve for $k^2=1$ shows oscillatory 
convergence. The solid blue curve for $k^2=0$ is also displayed for comparison.}}
\label{fig:GravDiscBessel}
\end{figure}

Recall that the series for the tachyon one-point function converges to the Bessel function 
$J_0$. Motivated by this, we rewrite the  expansion of ${\cal A}(-1)$ in terms of Bessel 
functions (using the Bessel expansion of $x^n$). This turns out to improve the behaviour 
of the series significantly. The graviton one-point function now shows oscillatory convergence 
(see Fig.\ref{fig:GravDiscBessel}) as in $\left<V\right>_D$. We should add, however, that 
this is also an asymptotic series and diverges for very large values of the argument (not
in the range of the graph).

We close by noting that the components of the energy-momentum tensor can be 
computed from these functions. However, in the presence of the linear dilation, a 
more careful analysis is needed as the Einstein metric differs from the string metric.

\sectiono{Summary and comments}\label{sec:Concl}
In this paper, we have considered a class of marginal deformations corresponding
to inhomogeneous decay of an unstable D-brane in the cubic open string field
theory. They satisfy the linearized equation of motion, and one of these, which
we call the Fisher deformation, gives a front solution that travels with a minimum
speed. We consider, in detail, the equations of OSFT to second order, and find that
the Fisher deformation also solves the equations to this order. A characteristic of
the marginal Fisher deformation is the appearance of a simple pole term in its OPE.
Thankfully, however, the operator that accompanies this singular term turns out to
be BRST-exact. This means that we are able to use the formalism developed for
marginal operators. It is likely that this deformation is exactly marginal, however,
the equations get rapidly very complicated and we leave the issue for future.

In the second part of our analysis, we computed the one-point functions of the
closed string tachyon (in particular, the disc partition function) and the gravitons
in the presence of the same marginal deformations on the boundary. These
expressions involve UV singularities corresponding to coincident operators on
the boundary. We have discussed one regularization scheme using Dyson
and related integrals found in the context of random matrix theory. We propose
a renormalisation that shows oscillations in the partition function before it decays
to the closed string vacuum. This feature, the physical consequences of which may
be worth exploring further, is in contrast to the case of homogeneous decay. 
There is also another spurious divergence at second order in the computation
of the graviton one-point functions.

Finally, the regularization and renormalization provide an exact expression for
the disc partition function in the background of the Fisher deformation. This may
allow one to use the formalism of background independent open string field
theory to get an alternative form of spacetime action for the tachyon field.

\bigskip

\noindent{\bf Acknowledgments:} It is a pleasure to thank Camillo Imbimbo, Dushyant
Kumar, Sunil Mukhi, Akhilesh Pandey, Ravi Prakash and Ashoke Sen. DG acknowledges
the audience of a NORDITA seminar for many interesting questions and comments, and 
Fawad Hassan and Konstantin Zarembo for hospitality. The work of DG was supported 
in part by SERC, DST (India) through the grant DST-SR/S2/HEP-043/2009, and PP is 
very grateful to Phra Jandee Jindatham and Watcharaporn Ladadok for inspiration and 
encouragement.

\end{document}